# Adam-Gibbs Model for the Supercooled Dynamics in OTP-OPP Mixture

C. M. Roland[1,*], S. Capaccioli[2], M. Lucchesi[2] and R. Casalini[1,3,†]

[1]Naval Research Laboratory, Chemistry Division, Code 6120, Washington DC 20375-5342

[2]INFM & Dipartimento di Fisica, Universitá di Pisa,Via Buonarroti 2, 56100 Pisa, Italy

[3]George Mason University, Chemistry Department, Fairfax, VA 22030

[*]*roland@nrl.navy.mil*          [†]*casalini@ccs.nrl.navy.mil*

**Abstract**  Dielectric measurements of the α-relaxation time were carried out on a mixture of *ortho*-terphenyl (OTP) with *ortho*-phenylphenol (OPP), over a range of temperatures at two pressures, 0.1 and 28.8 MPa. These are the same conditions for which heat capacity, thermal expansivity, and compressibility measurements were reported by Takahara et al. [S. Takahara, M. Ishikawa, O. Yamamuro, and T. Matsuo, Journal of Physical Chemistry B **103** (16), 3288 (1999).] for the same mixture. From the combined dynamic and thermodynamic data, we determine that density and temperature govern to an equivalent degree the variation of the relaxation times with temperature. Over the measured range, the dependence of the relaxation times on configurational entropy is in accord with the Adam-Gibbs model, and this dependence is invariant to pressure. Consistent with the implied connection between relaxation and thermodynamic properties, the kinetic and thermodynamic fragilities are found to have the same pressure independence. In comparing the relaxation properties of the mixture to those of neat OTP, density effects are stronger in the former, perhaps suggestive of less efficient packing.

## INTRODUCTION

The dramatic slowing down of molecular motions is one of the more intriguing phenomena accompanying the vitrification of liquids; however, the detailed physics underlying this behavior remains incompletely understood. As a liquid is cooled toward the glassy state, lower thermal energy hinders the ability of molecules to surmount the potential barriers on the energy landscape. The simultaneous thermal contraction promotes congestion and jamming, which also slow down molecular motions. Thus, in principle, temperature and density both have a role in governing the increase of relaxation times and viscosities during supercooling. Much experimental effort has been expended to quantify the relative contributions of temperature and density. Recent results indicate that both variables are important, and in the absence of specific interactions such as hydrogen bonding, they exert a roughly equivalent role. Scaling of experimental data based on an accounting of the density contribution have met with some success.[1-3]. However, models relying entirely on thermal activation or free volume to describe the supercooled dynamics cannot be correct.

Recent theoretical efforts have focused on the role of configuration entropy.[4-15] In the classic theory of Adam and Gibbs (AG)[16], the relaxation times of the supercooled liquid are determined by the configurational entropy, $S_c$, of the liquid according to

$$\tau = \tau_0 \exp\left(\frac{C_{AG}}{T S_c}\right) \tag{1}$$

in which $\tau_0$ and $C_{AG}$ are constants, the latter proportional to the free energy of activation for local rearrangements, $\Delta\mu$. The latter quantity is taken to be invariant to $T$ and $P$, although this assumption has been questioned.[17,18] The difficulty in applying the AG model is the unavailability of the configurational entropy. The original authors[16] and others since have assumed $S_c$ to be equal to an excess entropy, $\tilde{S}_e$, defined as the difference between the entropy of the liquid and the crystal

$$\tilde{S}_e = \int_0^T \left(C_{P,liq} - C_{P,cryst}\right) d\ln T \tag{2}$$

where $C_{P,liq}$ and $C_{P,cryst}$ are the isobaric heat capacities of the liquid and crystalline phases, respectively. The use of $\tilde{S}_e$ in place of $S_c$ is only an approximation, because the excess entropy includes the excess vibrational entropy, which has its own T-dependence.[19] From analysis of several liquids, $S_c \sim 0.7\,\tilde{S}_e$.[15,20] If $S_c$ is proportional to $\tilde{S}_e$, eq. 1 would still apply, with a renormalization of the constant $C_{AG}$.[21,20] However, the accuracy of this proportionality has been questioned.[19,22]

The problem can be circumvented by determining the entropy difference, $S_e$ between the liquid and the glass, since $S_e$ includes vibrational entropy [23]. The rapid rise in heat capacity just above $T_g$ is dominated by the strongly increasing configurational mobility. This means that the relatively weak T-dependence of the vibrational contributions to $S_e$ is negligible, at least over a limited temperature range around $T_g$. Accordingly, from the change with temperature in this region, the configurational entropy can be deduced, allowing a fairly accurate application of eq. 1.[24]

In this work we investigate the supercooled dynamics of a mixture of o-terphenyl (OTP) with o-phenylphenol (OPP). This liquid is of special interest because the heat capacity, thermal expansivity and compressibility have been measured at both ambient and elevated pressure.[25,26] OTP itself is a prototypical glass-former, studied by many techniques, including light scattering[27-31], neutron scattering[32-35], positronium annihilation spectroscopy[36], enthalpy and expansivity measurements[37-39], probe dynamics [40-42] and dielectric spectroscopy[43-45]. OTP is a fragile glass-former[46], whose relaxation times have a temperature dependence governed almost equally by temperature and density.[2,43] This strong influence of density enables the relaxation times for OTP to be expressed as a single function of the density[1,2]. Since OTP readily crystallizes, it is sometimes mixed with OPP in order to stabilize the supercooled state[26,37]. In the present experiments, we obtained isothermal dielectric relaxation spectra on mixtures of OTP with 33% by weight OPP. The measurements were made for various temperatures at ambient pressure and 28.8 MPa. These are the two pressures at which enthalpy and volumetric results were reported for the same composition.[25,26]. From the analysis, we evaluate the role of density, temperature, and configurational entropy in determining the variation of the dielectric α-relaxation times with temperature. We also assess the utility of the AG model in describing the data.

**EXPERIMENTAL**

OTP and OPP, obtained from Aldrich and used as received, were first mixed in their crystalline states at room temperature and then melted. The composition of the sample was 0.6691% OTP and 0.3309% OPP, by weight. Dielectric measurements, at both atmospheric and high (28.8 MPa) pressure, were carried out using a Novocontrol Alpha Analyzer ($10^{-2}$ to $10^7$ Hz). The sample was contained in a parallel plate capacitor (geometric capacitance ~ 10pF). For the measurements at atmospheric pressure, the sample was blanketed with nitrogen gas. For high pressure experiments, the sample was surround by silicon oil, and isolated from the pressurizing fluid by a Teflon seal. The dielectric cell was contained in a Cu-Be pressure vessel (UNIPRESS), with pressure applied using a manually operated pump (Nova Swiss). The pressure was measured with a Nowa Swiss tensometric transducer (0.1 MPa resolution). The temperature of the sample was monitored by a T-thermocouple in contact with the capacitor. Temperature was varied in the range 235–310 K by liquid flow from a thermostatic bath; stability was within 0.1 K.

**RESULTS AND DISCUSSION**

In Figure 1, we show a representative dielectric loss spectrum for the mixture at two temperatures for each pressure. The high pressure spectra have been shifted slightly to superimpose the peak maxima. It can be seen that when compared at values of $T$ and $P$ for which the relaxation times are equal, the shapes of the α-relaxation peaks are the same. We fit the peaks to the one-sided transform of the Kohlrausch-William-Watts function[47]

$$\phi(t) = \exp[-(t/\tau)^{\beta}] \qquad (3)$$

There is, some broadening with decreasing temperature, with $0.45 \leq \beta \leq 0.53$ over the range $-5 < \log \tau$ (s) $< -0.7$. For neat OTP, Naoki et al.[43] similarly found that the peak breadth at the same $\tau$ was pressure independent, but increased slightly with decreasing temperature, $\beta = 0.51 \pm 0.03$ for $-4.4 < \log \tau$ (s) $< -2.9$. Evidently, additional broadening of the α-dispersion due to the presence of 30% OPP is negligible.

When compared to neat OTP[48], the dielectric strength of the α-relaxation for the mixture is about an order of magnitude greater, due to the larger dipole moment of OPP. This means that the latter will contribute directly to the dielectric response, not only via its effect on the OTP dynamics. However, as described below, the characteristics of the relaxation properties, such as the Vogel-Fulcher parameters and the pressure coefficient of the glass temperature, for the mixture follow those of neat OTP, indicating that our measurements indeed probe the motion of the OTP.

In Figure 2 are shown the dielectric α-relaxation times, defined as the reciprocal of the peak frequency, $\tau = 1/(2\pi f_{peak})$, measured for OTP-OPP at both atmospheric pressure and 28.8 MPa. The range of the latter is limited, due to our inability to quench the sample into the supercooled state, because of the large thermal mass of our high pressure cell. (Without quenching, there is only a limited range of temperatures at which the supercooled liquid remains amorphous and homogeneous; at higher temperatures, some crystallization-induced phase separation appears to occur.) We fit the relaxation times to the Vogel-Fulcher (VF) equation

$$\tau = \tau_0 \exp\left(\frac{DT_0}{T - T_0}\right) \tag{4}$$

in which $D$ is a constant, and the Vogel temperature, $T_0$, can be identified with the Kauzmann Temperature, $T_K$, as has been shown for many liquids[49], including neat OTP[50,51]. In fitting isothermal relaxation measurements obtained at different pressures, we have previously shown that the parameter $D$ is independent of pressure.[52,53] Thus, we simultaneously fit the two data sets in Fig.2, using a common value of $D = 22.1 \pm 1$. The other best-fit parameters are listed in Table 1. Both $T$ and $\tau_0$ for the mixture are equal, to within the experimental error, to the values reported for neat OTP.[48] Note that the usual interpretation of the prefactor in eq. 4 is an attempt frequency, leading to the expectation $\tau_0 \sim 10^{-13}$s.[50] The values obtained from fitting the data in Fig. 2 are much shorter, too short to correspond to any physical process. The explanation for this lies in the failure of the VF function, when fitted to low temperature data, to describe relaxation times at high temperature, beyond some characteristic temperature, $T_B$. There is a change in dynamics at $T_B$, so that the value of $\tau$ calculated from eq. 4 cannot be extrapolated to high temperature.

Using the fitted VF, we obtain the temperature at which the relaxation time equals 1 s, $T_g = 241.3$ K and 247.2 K for $P = 0.1$ and 28.8 MPa, respectively. This corresponds to a pressure coefficient of the glass transition temperature equal to 0.206 K/MPa, which is significantly smaller than the value for neat OTP, $dT_g/dP = 0.260$.[54] These results are tabulated in Table 2.

A primary issue in analyzing data such as in Fig. 2 is the degree to which thermal energy and the density, $\rho$, govern the relaxation times, since both may contribute to a decreasing $\tau$ as temperature is reduced. In Figure 3a, we replot the isobaric data in Fig. 2 as a function of the density, using the published expansivity data for this mixture.[26]. Originally, Williams and coworkers[55,56] proposed the use of the ratio of the isochoric activation enthalpy, $H_V(T,V) = R \dfrac{\partial \ln \tau}{\partial T^{-1}}\Big|_V$ to the isobaric activation enthalpy, $H_P(T,P) = R \dfrac{\partial \ln \tau}{\partial T^{-1}}\Big|_P$ .The ratio varies from 0 and unity, reflecting an increasing dominance of temperature over density. The same information is contained in the ratio of the absolute value of the isochronal thermal expansion coefficient, $\alpha_\tau = -\rho^{-1}\left(\dfrac{\partial \rho}{\partial T}\right)_\tau$ to the isobaric thermal expansivity, $\alpha_P = -\rho^{-1}\left(\dfrac{\partial \rho}{\partial T}\right)_P$ .[57] These two quantities are related as[58]

$$-\alpha_\tau / \alpha_P = \left(\frac{H_P}{H_V} - 1\right)^{-1} \tag{5}$$

We have shown that this ratio is on the order of unit (or $H_V/H_P \sim 0.5$) for most van der Waals liquids near $T_g$ at low pressure[59]. To make this assessment for the OTP-OPP mixture, we use a relation due to Dreyfus et al.[2]

$$\frac{-\alpha_\tau}{\alpha_P} = \frac{\left(\dfrac{\partial \rho}{\partial P}\right)_T \left(\dfrac{\partial \ln \tau}{\partial T}\right)_P}{\left(\dfrac{\partial \rho}{\partial T}\right)_P \left(\dfrac{\partial \ln \tau}{\partial P}\right)_T} - 1 \tag{6}$$

At 248.7 K, which corresponds to $\tau = 0.01$ s at 0.1 MPa, we obtain $\left(\dfrac{\partial \rho}{\partial P}\right)_T / \left(\dfrac{\partial \rho}{\partial T}\right)_P = 0.452$ . From the data in Fig. 2, $\left(\dfrac{\partial \ln \tau}{\partial T}\right)_P = -0.566$ K$^{-1}$ at this temperature. The pressure coefficient in the denominator of eq. 6 is related to the activation volume, $\Delta V = RT\left(\dfrac{\partial \ln \tau}{\partial P}\right)_T$ . These are plotted in the inset to Fig. 2, from which we obtain $\left(\dfrac{\partial \ln \tau}{\partial P}\right)_T = -0.131 \pm 0.005$ MPa$^{-1}$ for $T = 248.7$ K. Eq. 6 yields $|\alpha_\tau|/\alpha_P = 0.96 \pm 0.07$ (or $H_V/H_P = 0.49 \pm 0.02$ from eq. 5). This is smaller than the value reported for neat OTP, $|\alpha_\tau|/\alpha_P \approx 1.3$ [2] ($H_V/H_P = 0.6 \pm 0.03$[43]), perhaps indicative of less efficient packing, and thus a stronger role for volume, in the mixture. The over-riding implication is that the variation of $\tau$ with temperature is due to density changes as much as to changes in thermal energy.

Tolle et al. [1] and Dreyfus et al.[2] were able to parameterize relaxation times for neat OTP, measured by neutron and light scattering respectively, using the quantity $T^{-1}\rho^{-4}$. We have recently demonstrated a more generalized scaling, $\log(\tau) \propto T^{-1}V^{-\gamma}$, which superimposes $\alpha$-relaxation times for a wide range of glass-forming liquids[3]. The scaling parameter $\gamma$ is material-specific, reflecting the relative contribution of volume to the temperature and pressure dependences. We obtained a master equation[3]

$$\frac{H_V}{H_P} = (1 + 0.19\gamma)^{-1} \qquad (7)$$

describing ten different liquids, of varying fragility.

In Fig. 3b, we show that, while the relaxation times measured for OTP-OPP at the two pressures are different, these $\tau$ are proportional to the product $T\rho^{-6.2}$. This value of $\gamma$, $= 6.2 \pm 0.3$, is consistent with eq. 7. $\gamma$ is larger for OTP-OPP than for neat OTP ($\gamma = 4$[1,2]) due to the stronger influence of density for the mixture. Thus, the scaling of the data in Fig. 3b is consistent with the magnitude activation enthalpy and expansivity ratios. These results are summarized in Table 2.

Although we can quantify the relative contribution of density and temperature to the relaxation behavior, entropy theories of the glass transition posit that the relaxation times should be a unique function of the configurational entropy, the latter subsuming the disparate effects of $\rho$ and $T$. The appeal of an entropy approach is that it provides a direct connection of the relaxation properties to thermodynamics, a connection which must, of course, exist.[60]

From the published heat capacities for the OTP-OPP, we calculate $S_e$, the excess entropy of the liquid over the glass. The more common excess entropy, $\tilde{S}_e$, defined with respect to the crystal entropy, is unavailable for this non-crystallizing mixture. Moreover, the entropy over the glass phase is a somewhat better estimate of the desired configurational entropy, since $S_e$ includes some of the excess vibrational entropy. The latter is not a part of the configurational entropy used in the AG equation, and thus is subtracted out in calculating $S_e$,

$$S_e = \int_0^T \left( C_{P,liq} - C_{P,glass} \right) d\ln T \qquad (8)$$

In eq. 8, $C_{P,glass}$ represents the isobaric heat capacity of the glass. The relaxation times for the mixture are plotted versus the reciprocal of the product $T \times S_e$ in Fig. 4. The data do not coincide, nor is either curve linear. This curvature demonstrates directly that the excess entropy cannot be used in place of the configuration entropy in applying eq. 1.

To calculate the configurational entropy, we fit the excess entropy above $T_g$ to a hyperbolic temperature dependence,[61].

$$S_e = a - \frac{b}{T} \qquad (9)$$

obtaining (Fig. 4 inset) $a = 125.0$ J/Kmol (for $P = 0.1$ MPa), and $a = 123.9$ J/Kmol ($P = 28.8$ MPa), with $b = 28.99$ kJmol[-1], independent of $P$. The rapid rise in heat capacity just above $T_g$ is dominated by the growth of configurational mobility, and over a limited temperature range above $T_g$, this effect dominates any changes in the vibrational entropy. Since $S_c$ differs from $S_e$ only by the exclusion in the latter of a small portion of the liquid configurational entropy, over this limited range, $S_c$ is expected to exhibit a temperature dependence similar to that of $S_e$; thus,

$$S_c = S_\infty - \frac{b}{T} \qquad (10)$$

where $S_\infty$ is the high temperature limiting value of the configurational entropy. Whereas $S_e$ goes to zero at $T_g$, $S_c = 0$ at the Kauzmann temperature. Since $T_K = T_0$ (specifically for neat OTP [50,51]), $S_\infty = b/T_K$. This gives

$$S_c = b \left( T_0^{-1} - T^{-1} \right) \qquad (11)$$

in which both parameters are known.

We are now in position to assess the AG model by plotting the relaxation times measured at the two pressures, according to the form of eq. 1. As can be seen in Figure 5, $\log \tau$ is directly proportional to $T^{-1} S_c^{-1}$, in conformance with the underlying assumption that the free energy of activation (potential barrier) for local rearrangements is independent of temperature. Moreover, the fact that the data for the two pressures are parallel implies that $\Delta\mu$ is also independent of pressure, at least up to P = 28.8 MPa. From the slope, we obtain $C_{AG}$ ($\propto \Delta\mu$) = 620 ± 2.4 kJ/mol. This is substantially larger than the value for neat OTP.[13]

Recently, a relationship was proposed between $S_c$ and the excess entropy of the melt over the crystal, $\tilde{S}_e$ [15]

$$S_c(T,P) = g_T(P_{atm})\tilde{S}_e(P_{atm}) - g_P(T)\int_{P_{atm}}^{P} \Delta\left(\frac{\partial V}{\partial T}\right)_{P'} dP' \qquad (12)$$

where $P_{atm}$ = 0.1 MPa, $g_T(P)$ and $g_P(T)$ are respective proportionality factors for the isobaric and isothermal components of $\tilde{S}_e$, and $\Delta\left(\dfrac{\partial V}{\partial T}\right)_P$ represents the difference between the expansivity of the liquid and that of the crystal. From eq.(12) an expression for the pressure-dependence of the Vogel temperature can be obtained[15]

$$T_0(P) = \frac{T_0(P_{atm})}{1 - \dfrac{g_P}{g_T}\dfrac{1}{S_\infty}\int_{P_0}^{P} \Delta\left(\dfrac{\partial V}{\partial T}\right)_{P'} dP'} \qquad (13)$$

where $S_\infty$ is evaluated at atmospheric pressure. Since the quantity $\Delta\left(\dfrac{\partial V}{\partial T}\right)_P$ is unknown for OTP-OPP, we calculate the difference between the thermal expansion coefficients for the liquid and glassy states from the data of Takahara et al.[26] These expansivities depend only weakly on temperature, and we obtain for the integral in the denominator of eq. 13, from 0.1 to 28.8 MPa, 3.31 ± 0.07 JK⁻¹mol⁻¹. Literature data for neat OTP suggest that the expansivity of the glass could be as much as 15% higher than $\left(\dfrac{\partial V}{\partial T}\right)_P$ for the crystal; thus, we take for the calculated value of the integral 3.5 ± 0.3 JK⁻¹mol⁻¹.

The two prefactors in eq. 12 account for the vibrational contribution to $\tilde{S}_e$. Since isobaric cooling affects both the density of states for the vibrational modes and the anharmonic potential, whereas isothermal compression affects only the former, we expect that $g_P(T) > g_T(P)$. This implies that $S_c$ is more efficiently reduced by an isothermal compression than by an isobaric compression, with consequently greater reduction in $\tau$ for the former. Since the values of $g_T$ and $g_P$ are unknown herein, we assume their ratio $g_T/g_P = 0.7$, as reported for other liquids including neat OTP.[15] Substituting these values into eq. 13, we obtain $T_0(28.8 \text{ MPa}) = 174.0 \pm 2.6$ K, which is in accord with the value determined directly from fitting the experimental relaxation times at the higher pressure (Fig. 2 and Table 1).

For different glass-formers, it has been found that application of pressure can cause the fragility, a measure of the departure of the relaxation times from Arrhenius behavior, to increase [53,62-64], decrease [9,52,65-68], or be invariant to pressure[58,69-72]. As shown in Fig. 6a, there is no change in the fragility of the OTP-OPP mixture, at least up to P = 28.8 MPa. We calculate the steepness index, $m \left( \equiv d \log(\tau) \big/ d \left( T_g/T \right) \big|_{T=T_g} \right) = 72 \pm 2$, which is in line with values determined from literature data for neat OTP[43,73],[39] $64 \leq \text{m} \leq 76$. An underlying idea of entropy models is that the rate of increase of the configurational entropy governs the non-Arrhenius behavior, whereby a correlation is expected between the fragility and the rate of change of the configurational entropy with temperature.[74] Various molecular liquids appear to conform to this idea,[74,75] although we have shown that the correlation fails for polymers[76-78]. In Fig. 6b the configurational entropy is plotted versus the reciprocal of temperature normalized by the Vogel temperature. The steepness of these curves is a measure of the thermodynamic fragility, and it can be seen that there is no effect of pressure. If a connection between thermodynamics and dynamics exists, this result is in accord with the invariance of *m* to pressure.

**Summary**

Dielectric relaxation measurements were obtained on a mixture of OTP with OPP (67/33) at both atmospheric and high (28.8 MPa) pressure. The data were analyzed by making use of previously published heat capacities and expansivities for the mixture at the same two pressures. The results are tabulated in Tables 1 and 2, and can be summarized as follows:

1. Both the glass transition temperature and its pressure coefficient are smaller for OTP-OPP in comparison to neat OTP. From this, it is tempting to infer that packing is less efficient in the mixture, due to the mismatch in size of the two molecules.

2. The α-peak in the dielectric loss, while broadening slightly with decreasing temperature, is independent of pressure. However, the breadth of the peak measured herein is equivalent to that for neat OTP; that is, there is no significant broadening due to concentration fluctuations.

3. Consistent with the invariance of the shape of the loss peak to pressure and to blending, the fragility of the mixture is independent of pressure, and equal to *m* for neat OTP.

4. An analysis of the *τ(T)* reveals that thermal energy and density exert an equivalent effect. The ratio of the isochronal and isobaric thermal expansion coefficients is *ca.* 20% smaller than the value of this ratio for neat OTP. This means density effects are augmented by blending.

5. The relaxation times for the mixture, as measured at various temperatures and two pressures, can be superimposed by expressing them as a function of the product $T\rho^{-6.2}$. This scaling parameter can be compared to value for neat OTP, $T\rho^{-4}$. The larger magnitude of the density exponent for OTP-OPP is consistent with the larger relative contribution of density to *T*-dependence of the relaxation times. The inference is that the poorer packing in the mixture emphasizes the effects of density.

6. To within the experimental error, the relaxation times of OTP-OPP for both pressures are a single, linear function of $T^{-1}S_c^{-1}$, indicating that the Adam-Gibbs model provides an adequate description of the supercooled dynamics over the modest range of temperatures herein. The results are consistent with an assessment of made using the expansivity data to calculate the change in the Vogel temperature with pressure. When plotted versus the Adam-Gibbs variable $T^{-1}S_c^{-1}$, the relaxation time data are independent of pressure, suggesting that the potential energy barrier for local rearrangements is invariant to pressure.

7. The steepness of the increase in the configurational entropy with temperature normalized by the Kauzmann temperature is independent of pressure. This is consistent with a connection between the kinetic and thermodynamic fragilities, since *m* is also invariant to pressure.

**Acknowledgements**

This work was supported by the Office of Naval Research.

Table 1. Results for OTP-OPP

| $P$ (MPa) | log $\tau_0$ (s) | $T_0$ (K) | $T_g(\tau = 1\text{s})$ (K) | $S_\infty$ (JK$^{-1}$mol$^{-1}$) |
|---|---|---|---|---|
| 0.1 | -21.7 ± 0.6 | 169 ± 2 | 241 | 172 ± 2 |
| 28.8 | -22.2 ± 0.7 | 174 ± 2 | 247 | 167 ± 2 |

Table 2. Comparison of OTP-OPP to neat OTP

| | $T_g$ (K)* | d$T_g$/d$P$ (K/MPa) | $\beta$ (eq. 3) | $M$ | $|\alpha_\tau|/\alpha_P$ | $\gamma$ |
|---|---|---|---|---|---|---|
| neat OTP | 247[54] | 0.260[54] | 0.51 ± 0.03[43] | 70 ± 6[43,73,39] | 1.3[2] | 4[1,2] |
| OTP-OPP | 233 | 0.206 | 0.49 ± 0.04 | 72 ± 2 | 0.96 ± 0.07 | 6.2 ± 0.3 |

[*] thermal analysis

**Figure Captions**

Figure 1. α-dispersion in the dielectric loss for OTP-OPP measured at 0.1 MPa (solid symbols) and 28.8 MPa (hollow symbols) and T = 249.4 (●), 256.3 (▲,□) and 262.7 (▽) K. The higher pressure spectra have been shifted to superimpose the peak maxima: horizontally by 0.6 and vertically by 1.05 (256.3K); horizontally by 0.9 and vertically by 1.11 (262.7K). The solid lines are the fits to the transform of eq. 3 with the indicated value of the Kohlrausch exponent.

Figure 2. α-relaxation times measured for OTP-OPP at the indicated pressures, along with the fitted VF curves (eq.4), with $D$ = 22.1 ± 1 and the other parameters given in Table 1. The inset shows the activation volume calculated at each temperature for which measurements at 28.8 MPa were made.

Figure 3. α-relaxation times at the indicated pressures as a function of (a) the mass density and (b) the product of the temperature times the density to the -6.2 power.

Figure 4. α-relaxation times for OTP-OPP plotted versus the excess entropy in the manner suggested by the AG equation. The error bars are smaller than the symbol size. $S_e$ was determined from heat capacity measurements at ambient and elevated pressures[25], and is plotted in the inset above the calorimetric $T_g$ = 233.7 K.

Figure 5. α-relaxation times for OTP-OPP plotted versus the configurational entropy in the manner suggested by the AG equation. The differences between the data for the two pressures are within the error bars.

Figure 6. (a) α-relaxation times as a function of the inverse of the temperature normalized by the temperature at which $\tau$ = 1 s. (b) configurational entropy as a function of the inverse of the temperature normalized by the Vogel temperature. In both figures, the solid circles are for $P$ = 0.1 MPa and the open squares for $P$ = 28.8 MPa.

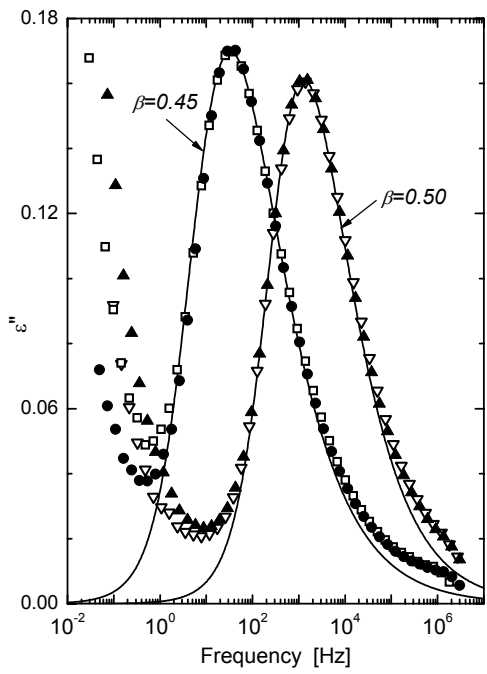

figure 1

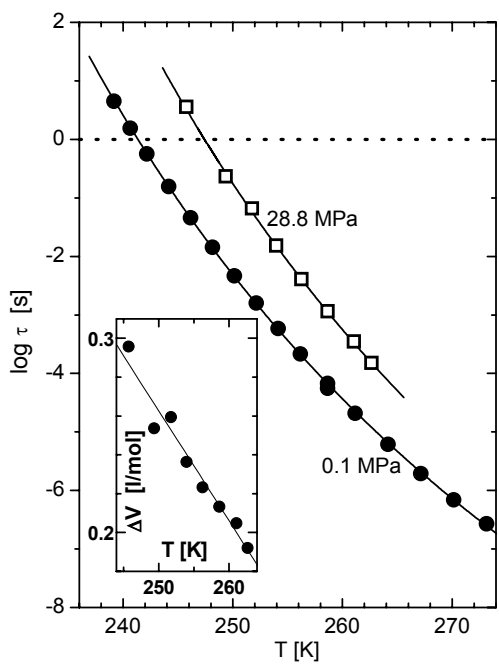

figure 2

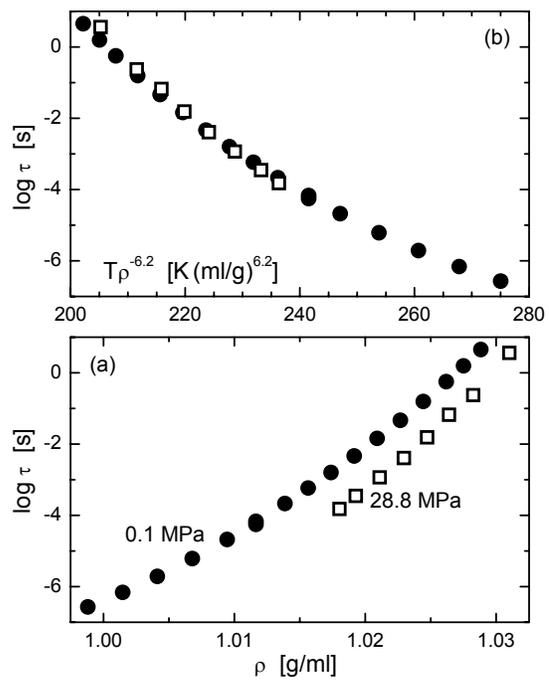

figure 3

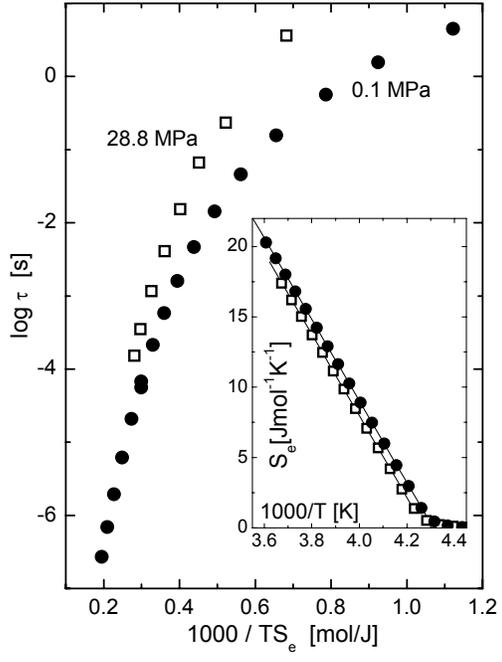

figure 4

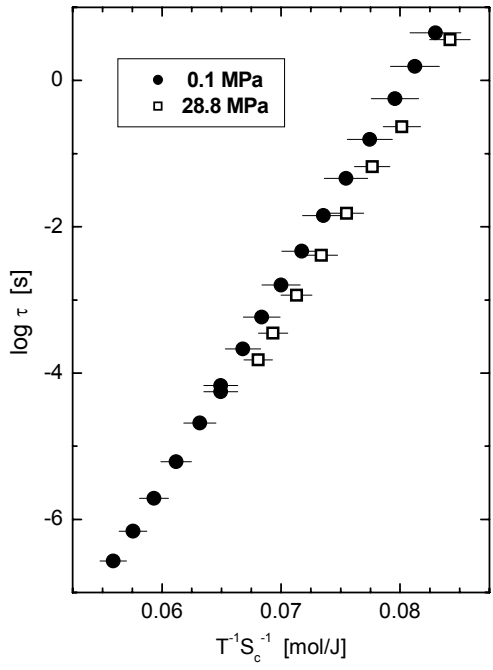

figure 5

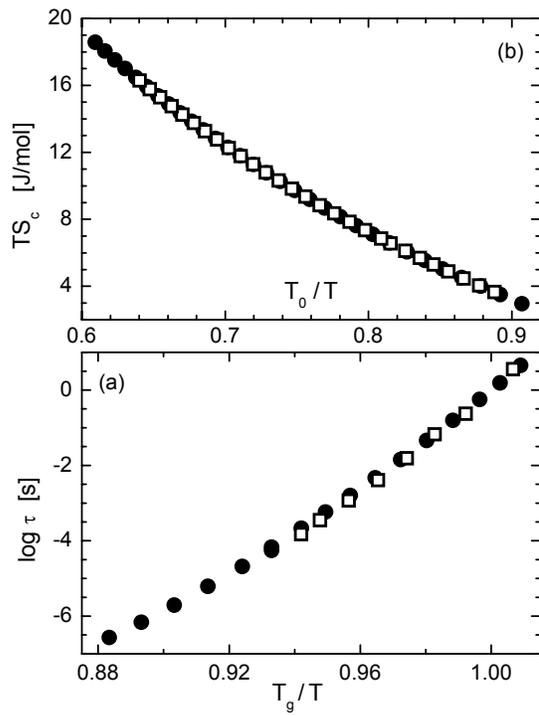

figure 6